\documentclass[prX,twocolumn]{revtex4}
\usepackage[T1]{fontenc}
\usepackage{epsfig}
\newcommand{\ud}{\mathrm{d}}
\begin{document}
\title{%
The WIMP capture process for dark stars in the early universe}
\author{Sofia Sivertsson}
\email{sofiasi@kth.se}
\affiliation{Department of Theoretical Physics, Royal Institute of Technology (KTH), AlbaNova University Center, 106 91 Stockholm, Sweden and \\
The Oskar Klein Centre for Cosmoparticle Physics, Stockholm University, AlbaNova University Center, 106 91 Stockholm, Sweden}
\author{Paolo Gondolo}
\email{paolo.gondolo@utah.edu}
\affiliation{Department of Physics and Astronomy, University of Utah, 115 South 1400 East \#201, Salt Lake City, Utah 84112-0830, USA}
\date{May 31, 2010}
\begin{abstract}
\noindent The first stars to form in the universe may have been dark stars, powered by dark matter annihilation instead of nuclear fusion. The initial amount of dark matter gathered by the star gravitationally can sustain it only for a limited period of time. It has been suggested that capture of additional dark matter from the environment can prolong the dark star phase even to the present day. Here we show that this capture process is ineffective to prolong the life of the first generation of dark stars. We construct a Monte-Carlo simulation that follows each Weakly Interacting Massive Particle (WIMP) in the dark matter halo as its orbit responds to the formation and evolution of the dark star, as it scatters off the star's nuclei, and as it annihilates inside the star. A rapid depletion of the WIMPs on orbits that cross the star causes the demise of the first generation of dark stars. We suggest that a second generation of dark stars may in principle survive much longer through capture. We comment on the effect of relaxing our assumptions.
\end{abstract}
\maketitle

The first stars to form in the universe (Population III stars) provide the light that ends the dark ages, contribute to reionization, inject metals into the interstellar medium for later stellar generations, and might give the seeds for the massive early black holes observed in quasars. All these effects depend on the properties of the first stars, which in turn depend strongly on their masses. 

Dark matter can play a crucial role in the formation of the first stars \cite{sfg}, and in particular can change their masses and astrophysical implications dramatically \cite{Spolyar:2009nt}. The first stars form at the centers of early dark matter halos, when the universe was young and much more dense than today. Not only were the first stars formed in regions of very high dark matter densities, but they also dragged in more dark matter gravitationally while they were forming (a process often called ``adiabatic contraction'' although it may not occur adiabatically). If dark matter consists of Weakly Interacting Massive Particles (WIMPs) that annihilate among themselves, their annihilation could provide enough energy to stop or delay the formation of Pop III stars, giving rise to a new stellar phase called a Dark Star. This annihilation is the same physical process that creates WIMPs thermally  in the early universe and naturally gives the correct cosmic density for WIMPs to account for the observed dark matter. 

Once the initial amount of dark matter accumulated via adiabatic contraction runs out, the dark star phase ends and the star contracts onto the zero-age zero-metallicity main sequence \cite{Spolyar:2009nt}. It has been suggested \cite{Yoon:2008km,Taoso:2008kw,Freese:2008ur} that capture of additional dark matter from the environment around the star can prolong the dark star phase even to the present day. 

Here we show that this capture process is ineffective to prolong the life of the first generation of dark stars. For these stars, the great majority of the dark matter around them is accumulated by adiabatic contraction at the time of formation and ends up in orbits that are bound to the dark star. Part of this dark matter is captured by the dark star once the gas density in the star is high enough. Essentially all of the dark matter in the dark star annihilates away during its contraction towards the main sequence at the end of the dark star phase. The rest of the dark matter remains close but outside the star and orbits around it without crossing it.  We call this dark matter population the dark matter envelope. Once the dark matter crossing the star has been depleted, there is no additional dark matter population that can be efficiently captured. This ends the first generation of dark stars.

Notice that in this paper we have extended the meaning of the word capture in the following sense. By capture, we mean the concept of bringing WIMPs from ``outside'' to ``inside'' the star through energy loss in WIMP-nucleon scattering, which increases the fraction of time a WIMP spends inside the star. Originally capture was defined as the scattering of dark matter particles in the Galactic population off individual nucleons in the Sun and into bound orbits \cite{1985ApJ...296..679P,Gould:1987ir}. In the case of Pop III dark stars, most of the WIMPs accumulated around the star by adiabatic contraction are bound to the star, and there is no relevant unbound population. Hence the original formalism for capture, as used in~\cite{Yoon:2008km,Taoso:2008kw,Freese:2008ur}, does not apply. We instead define capture as the scattering of dark matter particles off nucleons in the star without regard to the bound or unbound nature of their orbits. This definition allows for a clear separation of scattering and annihilation processes.

\section{Introduction}

WIMPs constitute a negligible fraction of the mass of the dark star (less than a few percent), but the WIMP density is high enough for WIMP annihilation to become the dominant power source. This was proposed in Ref.~\cite{sfg}, which introduced the term Dark Stars to refer to Pop III stars in a stellar evolution phase powered by dark matter. Later the term was extended in Ref.~\cite{Scott:2008ns} to any star powered by dark matter annihilation, previously called a dark matter burner \cite{Moskalenko:2006mk,Moskalenko:2007ak,Bertone:2007ae,Fairbairn:2007bn}. Work on Pop III dark stars include Refs.~\cite{sfg,Iocco:2008xb,Freese:2008ur,Freese:2008hb,Iocco:2008rb,Freese:2008wh,Yoon:2008km,Taoso:2008kw,Freese:2008ct,Freese:2009ua,Spolyar:2009nt,Ripamonti:2009xw,Freese:2010re,Ripamonti:2010ab}; work on other kinds of dark stars include Refs.~\cite{Moskalenko:2006mk,Moskalenko:2007ak,Bertone:2007ae,Fairbairn:2007bn,Scott:2007md,Scott:2008ns,Scott:2008uw,Scott:2009nd,Casanellas:2009dp}. In this paper, we consider Pop III dark stars exclusively.

In dark stars, the dark matter becomes the dominant heat source despite its comparably low density because WIMP annihilation converts mass into energy much more efficiently than nuclear fusion. As in Ref.~\cite{sfg}, we assume that $2/3$ of the WIMP annihilation energy is deposited as heat in the star and the other $1/3$ is lost, predominantly to neutrinos. 

As the dark star forms, the deepening of the gravitational potential boosts the halo's central WIMP density in the proximity of the star (adiabatic contraction). Annihilation within the WIMP population heats up the dark star, but also decreases the dark matter density. However, as long as the dark star keeps gaining mass and increases its density, the central dark matter population inside the star is replenished by the increased gravitational pull. The energy injected by WIMP annihilation halts the contraction of the star, keeping the outer regions cool for a time longer than a normal Pop III star without dark matter. The dark star is then supported by the energy from the annihilation of WIMPs. Initially, it is a low-density giant star with low surface temperature \cite{sfg}. Later, by accreting mass from the surrounding gas, it grows to a mass of several hundreds of solar masses, still remaining a low-density giant star (radius of thousands of solar radii). The dark matter has given the star more time to accrete mass, making the first stars more massive than without the help from dark matter annihilation.

As the dark star becomes bigger and more massive, more energy is needed to support it. To sustain the star, dark matter needs to be supplied at a higher rate. Eventually the adiabatic contraction can no longer keep up with the dark star's increasing demand of dark matter. At this point, the dark star starts to contract since the dark matter supply rate is no longer enough to support it. When the volume of the dark star decreases, the total amount of energy supplied by dark matter annihilation decreases with the volume. After the contraction, the star approaches the main sequence. When nuclear fusion starts, the radiation feedback halts the accretion of baryons, putting an end to the star's growth in mass. Then the dark matter can no longer be gravitationally replenished by adiabatic contraction, putting an end to the initial dark star phase.
The end product is a normal, although very massive, main sequence star in the sense that it produces energy through nuclear fusion and no longer through WIMP annihilation.

It was suggested in Refs.~\cite{Yoon:2008km,Taoso:2008kw,Freese:2008ur} that the WIMP population at the end of the dark star phase may be replenished by capture of WIMPs from the surrounding dark matter halo. WIMPs passing through the star can loose energy in scattering off nucleons in the star. If a WIMP scatters many times it will eventually loose enough energy to sink to the center and contribute to the WIMP density inside the star. If enough energy is injected through the annihilation of the captured dark matter, the nuclear burning in the star is slowed down, possibly to essentially a full stop. This object would then be powered by dark matter and, as long as the dark matter fuel is constantly refilled, the lifetime of the star is substantially prolonged, possibly to time scales comparable to the age of the universe (``eternal dark stars'' \cite{Yoon:2008km,Taoso:2008kw,Freese:2008ur}).

For WIMP capture to prolong the life of a dark star, it is essential that the dark matter is efficiently captured via WIMP-nucleon scattering. Refs.~\cite{Yoon:2008km,Taoso:2008kw,Freese:2008ur} estimated the rate of WIMP capture using a formula by Gould \cite{Gould:1987ir}. However, Gould's formula assumes a constant flow of WIMPs coming onto the star from far away. It also assumes that the WIMP population is Maxwellian and is not bound to the star. However, a Pop III dark star forms thanks to the adiabatic contraction of its own dark matter halo, and the WIMPs responsible for the high concentration of dark matter in and around the star are gravitationally bound to the star. Thus also the WIMPs accessible for capture are gravitationally bound to the star, and Gould's formula does not apply. 

In this work, we investigate in detail the star's ability to access dark matter from the halo it formed in. For this purpose, we construct a Monte-Carlo simulation that follows the halo WIMPs as their orbits respond to the formation of the star, as they scatter off the star's nuclei, and as they annihilate inside the star. Since the WIMPs are fully traced, not only can we analyze their annihilation and scattering behavior, but also their phase space distribution around the formed star.

We find that WIMP capture is not important until the dark star contracts on its way to the main sequence and becomes dense enough to be a viable dark matter target. We also find that WIMP capture lasts a relatively short amount of time as the WIMPs available for capture are quickly depleted and their orbits are not replenished.

Throughout this paper we use as fiducial values a WIMP mass $m_\chi =100$~GeV and an annihilation cross section times relative velocity $\sigma_{\rm ann} v =3\times 10^{-26}$~cm$^3/$s. For scattering off nuclei, since the spin dependent scattering cross section is far less constrained than the spin independent cross section, we look at the case of a large spin-dependent scattering cross section $\sigma_{\rm scatt}=10^{-39}$~cm$^2$, as well as at a zero scattering cross section (absence of scattering). 

Section II describes our Monte-Carlo (initial phase space distribution, adiabatic contraction, capture and annihilation). There we also present a new quadrature formula for the phase-space distribution of a Navarro-Frenk-White \cite{NFW} halo. Section III presents our results, which are then discussed in Section IV.

\section{Method}

The motivation for this work is to find out how the dark matter in the NFW halo responds to the change in gravitational potential as the star forms, and what then happens to the dark matter with time. Following the choices of Ref.~\cite{sfg} we investigate the case of a single star forming at redshift $z=20$ in the center of a primordial halo of mass $10^6 M_\odot$ with a Navarro, Frenk and White (NFW) profile \cite{NFW} consisting of 15\% baryons and 85\% dark matter.

The Monte-Carlo simulation picks individual WIMPs from the original NFW halo and follows these WIMPs and how they respond to the changing gravitational potential as the star forms. Through picking and studying WIMPs in phase space the full orbit of the WIMP investigated is known. With this information the adiabatic invariants of the orbit can be determined and with that, as long as the star formation process is slow enough to be adiabatic, the new orbits of the WIMPs can be found through all stages of the stellar evolution. The phase space distribution of the dark matter as the star evolves is thus followed. 

WIMPs spending time inside the dark star can annihilate with other WIMPs or they can scatter off nuclei in the star. If the WIMPs annihilate, their energy (minus the energy lost to neutrinos) is released to heat the gas. If the WIMPs scatter before having time to annihilate, the energy loss and then the new WIMP orbit is calculated. The new orbit of the scattered WIMP is then followed and the WIMP is again given the chance to scatter or annihilate. As the star evolves, the future orbits of the WIMPs which have not annihilated are calculated, so that the WIMPs are given new chances to scatter and/or annihilate at the later stages of the stellar evolution. Simulating many WIMPs gives the injected heat from dark matter into the dark star as a function of time, including the effect of WIMP-nucleon scattering. 

The evolution used for the dark star is taken from Ref.~\cite{Spolyar:2009nt}. More precisely, it is their canonical case of a 100 GeV WIMP with Tan-McKee mass accretion  \cite{Tan:2003bs} and without WIMP capture (i.e.\ without WIMP-nucleon scattering). To follow the WIMPs through the formation of the star, the stellar evolution is divided into 116 steps (taken from \cite{Spolyar:2009nt}) and the WIMP orbits are found for each of these stages. Our work only considers how the dark matter halo evolves given the stellar evolution. How the dark star responds to the change of the dark matter heating from WIMP capture will be taken into account in an upcoming publication \cite{myself}.

\subsection{Initial dark matter distribution}
The first step of the Monte-Carlo is to pick WIMPs from the original NFW halo, i.e.\ to find the phase space distribution of the original halo. Spherical symmetry is assumed to hold throughout, for the star forming in the center of the NFW halo and at later times.

The density profile of an NFW halo is given by 
\begin{equation}
\frac{\rho(r)}{\rho_{c}}=\frac{\delta_c}{(r/r_s)(1+r/r_s)^2},
\end{equation}
where $r_s=r_{200}/c$ is a characteristic radius and $\rho_{c}=3H^2/8\pi G$ is the critical density. $\delta_c$ is a dimensionless concentration parameter given by the requirement that the mean density within $r_{200}$ is given by $200\rho_{c}$, as in Ref.~\cite{NFW}. We take the total halo mass to be $10^6\, M_\odot$ and the concentration parameter $c=2$. Then $\delta_c=1235$, $r_s=2.28\times10^{20}\,{\rm cm}=74.5\,{\rm pc}$, and $\rho_{c}=2.5\times10^{-26}\,{\rm g/cm^3}=0.014\,{\rm GeV/cm^3}$.

For a spherically symmetric density distribution and an everywhere isotropic velocity dispersion, the phase-space distribution function can be found using Eddington's formula \cite{Binney}:
\begin{equation}
f(\mathcal E)=\frac{1}{\sqrt{8}\pi^2}\left[\int^{\mathcal E}_0 \frac{\ud\Psi}{\sqrt{\mathcal E-\Psi}}\frac{\ud^2\rho}{\ud\Psi^2}+\frac{1}{\sqrt{\mathcal E}}\left(\frac{\ud\rho}{\ud\Psi}\right)_{\Psi=0}\right].
\end{equation}
Here $f({\mathcal E})$ has units of mass per unit phase space volume (g cm$^{-3}$ (cm/s)$^{-3}$), $\mathcal E=-E/m_\chi$ is the negative of the total energy $E$ per unit WIMP mass $m_\chi$, and $\Psi(r)=-\phi(r)$ is the negative of the gravitational potential $\phi(r)$. The condition that the WIMPs are gravitationally bound to the NFW halo implies $0<\mathcal E\leq \Psi(0)$. For an NFW halo, $\Psi(r)$ is given by:
\begin{equation}
\Psi_{\rm NFW}(r)=G\int_r^\infty\frac{M(r')}{r'^2}\ud r'=4\pi G\delta_c\rho_{c}r_s^2\frac{\ln(r/r_s+1)}{r/r_s}.
\end{equation}
We have written $M(r)$ for the halo mass enclosed within radius $r$.

\begin{figure}
\centerline{\epsfig{file=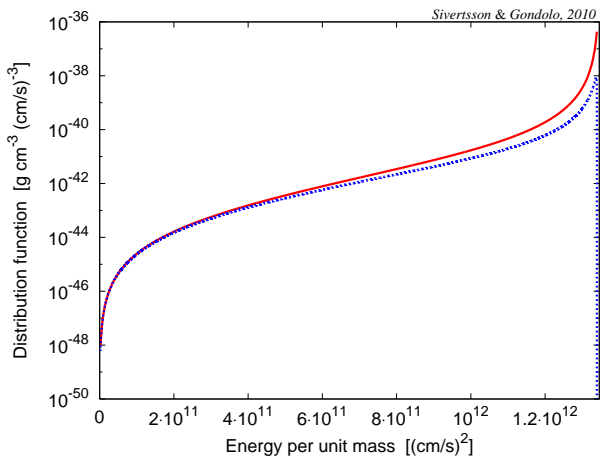,width=0.48\textwidth}}
\caption{The phase-space distribution function $f_{\rm NFW}({\mathcal E_{\rm NFW}})$ (red solid line) and the ${\mathcal E_{\rm NFW}}$-distribution $f_{\rm NFW}(\mathcal E_{\rm NFW})\sqrt{1-{\mathcal E_{\rm NFW}}/\Psi_{\rm NFW}(r)}$ at $r=10^{-2}r_s$ (blue dotted line) for our $10^6\,M_\odot$, $c=2$ NFW halo. }
\label{fig:fE}
\end{figure}

After a very lengthy but straightforward calculation starting from Eddington's formula, the distribution function for an NFW profile can be written as:
\begin{equation}
f_{\rm NFW}(\mathcal E_{\rm NFW})=\frac{1}{\sqrt{8}\pi^2}\int_{r_{\mathcal E^{\rm NFW}}}^\infty g(r) \frac{1}{\sqrt{\mathcal E_{\rm NFW}-\Psi_{\rm NFW}(r)}}\ud r,
\label{eq:fE}
\end{equation}
where
\begin{equation}
g(r)=\frac{r(r_s-3r+6\xi)}{4G\pi(r_s+r)^3(r-\xi)^2}
\end{equation}
with
\begin{equation}
 \xi=(r_s+r)\ln\left(1+\frac{r}{r_s}\right).
\end{equation}
Moreover, the lower limit of integration in Eq.~(\ref{eq:fE}), $r_\mathcal E^{\rm NFW}$, is the solution of $\Psi_{\rm NFW}(r_\mathcal E^{\rm NFW})=\mathcal E_{\rm NFW}$. This solution cannot be written in compact form. 
The NFW phase-space distribution $f_{\rm NFW}({\mathcal E_{\rm NFW}})$ is plotted in Fig.~\ref{fig:fE}.

Knowing $f_{\rm NFW}({\mathcal E}_{\rm NFW})$, the Monte-Carlo can pick individual WIMPs from the original halo. In practice, we first pick a spatial location $r$ from the distribution $4\pi r^2\rho(r)\ud r$, then we select a WIMP energy from the distribution $4\pi f_{\rm NFW}(\mathcal E_{\rm NFW})\sqrt{2[\Psi_{\rm NFW}(r)-\mathcal E_{\rm NFW}]}\ud \mathcal E_{\rm NFW}$, and finally we generate a direction for the WIMP velocity from a uniform distribution on the sphere. The latter two distributions use the isotropy in velocity space. The WIMP angular momentum per unit mass $\mathcal J$ is easily determined from ${\mathcal E_{\rm NFW}}$, $\Psi_{\rm NFW}(r)$ and the velocity direction at the point $r$: $\mathcal J=r\sin\theta\sqrt{2[\Psi_{\rm NFW}(r)-\mathcal E_{\rm NFW}]}$, where $\theta$ is the angle between the velocity vector and the outward radial direction. With this information, the WIMP's orbit is completely determined up to the spherical symmetry. 

There is a solvable numerical difficulty in generating ${\mathcal E_{\rm NFW}}$, connected to the divergence of $f_{\rm NFW}({\mathcal E_{\rm NFW}})$ as ${\mathcal E_{\rm NFW}} \to \Psi_{\rm NFW}(0)$. The divergence is in itself not a problem because a finite radius $r$ of the WIMP location avoids it as $\mathcal E_{\rm NFW}\leq\Psi_{\rm NFW}(r)<\Psi_{\rm NFW}(0)$; thus in practice the divergence point is never reached since $M(r)\to0$ as $r\to0$. The finite value of $r$ guarantees that the distribution from which the WIMPs are generated, namely $4\pi f_{\rm NFW}(\mathcal E_{\rm NFW})\sqrt{2[\Psi_{\rm NFW}(r)-\mathcal E_{\rm NFW}]}\ud \mathcal E_{\rm NFW}$, remains finite everywhere. This distribution is shown in Fig.~\ref{fig:fE} (blue dotted line), rescaled to fit $f(\mathcal E)$ for small $\mathcal E$. Even though the distribution from which the WIMPs are picked is finite, the divergence in $f(\mathcal E)$ still produces a sharp peak near ${\mathcal E_{\rm NFW}}=\Psi_{\rm NFW}(r)$, as seen in Fig.~\ref{fig:fE}. In the Monte-Carlo, the peak is typically even sharper than in Fig.~\ref{fig:fE}, as $r$ is typically smaller for the interesting WIMPs, making the distribution numerically challenging to use in the Monte-Carlo. The latter issue can be treated by going back to Eq.~(\ref{eq:fE}); Taylor expanding around the critical point ${\mathcal E_{\rm NFW}}=\Psi_{\rm NFW}(r)$, and integrating in $r$, one finds that the peak can be tamed by rewriting the velocity distribution in the new variable $z=1/(\Psi_{\rm NFW}(0)-\mathcal E_{\rm NFW})$. 

\subsection{Adiabatic contraction}

As the star forms, the gravitational potential changes from $\Psi_{\rm NFW}(r)$ to $\Psi_\star(r) + \Psi_{\rm halo}(r)+\Psi_{\rm gas}(r)$, where $\Psi_\star(r)$ is the potential generated by the star, $\Psi_{\rm halo}(r)$ is the potential generated by the dark matter halo while it adiabatically contracts, and $\Psi_{\rm gas}(r)$ is the potential generated by the gas surrounding the star.

The star influences the potential out to a radius of influence $r_{\rm infl}$, which can be estimated by equating the mass of the star $M_\star$ to the amount of dark matter in the halo contained within $r_{\rm infl}$, i.e.~setting $M_{\rm halo}(r_{\rm infl})=M_\star$. To find $M_{\rm halo}(r_{\rm infl})$ one can use adiabatic contraction in the circular orbit approximation, in which WIMP orbits do not cross each other so that
$
M_{\rm NFW}(r_{\rm infl}^{\rm MFW})=M_{\rm halo}(r_{\rm infl}) .
$
Here $r_{\rm infl}^{\rm MFW}$ is the radius the WIMP orbit had before it adiabatically contracted to $r_{\rm infl}$. Using this relation with the NFW density profile we obtain $r_{\rm infl}^{\rm NFW} \simeq \sqrt{M_\star /(2 \pi \rho_c \delta_c r_s)} \simeq 0.025 r_s \sqrt{M_\star/800M_\odot}$. For circular orbits, $rM(r)$ is an adiabatic invariant. After the formation of the dark star, the mass doubles inside a WIMP orbit originating at $r_{\rm infl}^{\rm MFW}$ and the invariance of $rM(r)$ then tells us that $r_{\rm infl}=r_{\rm infl}^{\rm NFW}/2\simeq 0.01 r_s \sqrt{M_\star/800M_\odot}$.
Another way to estimate the radius of influence is to estimate how far a WIMP would go if it would leave the surface of the star at a speed equal to the star's escape velocity $v_{\rm esc} = \sqrt{2GM_\star/R_\star}$. WIMPs which travel further out in the halo are not gravitationally bound to the star. For the density distribution of the original NFW halo, one finds in this way that $r_{\rm infl}^{\rm NFW} \simeq \sqrt{M_\star /(2 \pi \rho_c \delta_c r_s)}$, which is the same expression as in the previous estimate of the radius of influence. In conclusion, a good estimate of the radius of influence is $r_{\rm infl} \sim 0.01 r_s$ for a final star of mass $M_\star = 800 M_\odot$ and radius $R_\star=4\times10^{11}\,{\rm cm}$.

Within the star and its immediate surroundings, the star contribution $\Psi_\star(r)$ dominates over $\Psi_{\rm halo}(r)$, and thus in that region we neglect $\Psi_{\rm halo}(r)$. This approximation is good out to the radius of influence of the star $r_{\rm infl}$.  By neglecting the halo component $\Psi_{\rm halo}(r)$, we have implicitly assumed that all WIMPs close to the star are bounded to the star. In Section IV we show that the contribution of unbound WIMPs is negligible.

Similarly, the density of the gas surrounding the star is so low that the potential $\Psi_{\rm gas}(r)$ it generates can be neglected. For example, for a mass accretion rate $\dot{M} = 10^{-3} \, M_\odot/{\rm yr}$, in the middle of the range of the Tan-McKee \cite{Tan:2003bs} and O'Shea-Norman \cite{O'Shea:2006tu} accretion rates, the hydrodynamic equation gives a gas density that falls as $r^{-3/2}$ from the surface of the star, where it is approximately $\rho_{\rm gas}(R_\star) = \dot{M}/(4\pi R_\star \sqrt{GM_\star R_\star}) \sim 10^{-10}\,{\rm g/cm^3}$. This is to be compared with an average stellar density of $\sim 6 \, {\rm g/cm^3}$ for the final stages of the dark star ($M_\star = 800 M_\odot$, $R_\star=4\times10^{11}\,{\rm cm}$). As a consequence, the gas mass $M_{\rm gas}(r)$ inside a radius $r$ is roughly one half of the stellar mass $M_\star$ for $r=r_{\rm infl}$. Hence also $\Psi_{\rm gas}(r)$ can be neglected for the region $r<r_{\rm infl}$ in which we are interested.

Within the star and its immediate surroundings, the star contribution $\Psi_\star(r)$ dominates over both $\Psi_{\rm halo}(r)$ and $\Psi_{\rm gas}(r)$, and thus in that region we neglect $\Psi_{\rm halo}(r)$ and $\Psi_{\rm gas}(r)$. This approximation is good out to the radius of influence of the star $r_{\rm infl}$.  By neglecting the halo components $\Psi_{\rm halo}(r)$ and $\Psi_{\rm gas}(r)$, we have implicitly assumed that all WIMPs close to the star are bound to the star. In Section IV we show that the contribution of unbound WIMPs is negligible. This tells us that we can safely assume that the potential in the region of interest is dominated by the potential of the star $\Psi_\star(r)$. The potential inside the star can be obtained numerically from the tabulated dark star evolution models in Ref.~\cite{Spolyar:2009nt}. Outside the star, the potential $\Psi_\star(r)$ is simply Keplerian. 

The individual WIMP's response to the adiabatic contraction of the forming star can be followed using the adiabatic invariants
\begin{equation}
\mathcal J \mbox{ \ \ and \ \ } \mathcal J_r=2\int_{r_{\min}}^{r_{\max}}\sqrt{2[\Psi(r)-\mathcal E]-\mathcal J^2/r^2}\,\ud r,\label{adiabatic_invariants}
\end{equation}
where $\mathcal J$ is the angular momentum per unit WIMP mass and ${\mathcal J}_r$ is the radial action per unit WIMP mass.
These invariants assume the same value before, during, and after the formation of the star, as long as the WIMPs do not scatter. This allows us to generate the distribution of WIMPs as they enter the star for the first time.

The energy ${\mathcal E}$ is not an adiabatic invariant, however, and needs to be recomputed as the potential changes. For this, we first compute ${\mathcal J}_r$ for the NFW potential using the energy $\mathcal E_{\rm NFW}$ generated from $f_{\rm NFW}(E_{\rm NFW})$ as explained above:
\begin{equation}
\mathcal J_r^{\rm NFW}=2\int_{r^{\rm NFW}_{\min}}^{r^{\rm NFW}_{\max}}\sqrt{2[\Psi_{\rm NFW}(r)-\mathcal E_{\rm NFW}]-\mathcal J^2/r^2}\,\ud r .
\end{equation}
As $\mathcal J_r$ is an adiabatic invariant $\mathcal J_r=\mathcal J_r^{\rm NFW}$. We then solve the following equation for the new energy $\mathcal E$:
\begin{equation}
2\int_{r_{\min}}^{r_{\max}}\sqrt{2[\Psi_\star(r)-\mathcal E]-\mathcal J^2/r^2}\,\ud r = \mathcal J_r^{\rm NFW}  \label{eq:Jr-star}.
\end{equation}
Here the limits of integration  $r_{\min}$ and $r_{\max}$ are the zeros of the square root. In writing this equation, we have assumed that $\mathcal E>0$, consistent with the fact that WIMPs close to the star are bound to it.

For WIMPs on orbits that do not cross the star, the radial action equation can be solved analytically since their orbits are Keplerian. The solution is \cite{goldstein}
\begin{equation}
\mathcal E = \frac{2\pi^2 G^2M_\star^2}{(\mathcal J_r+2\pi\mathcal J)^2} .
\label{eq:EKepler}
\end{equation}

For WIMPs that cross the star the orbit is not Keplerian and $\mathcal E$ needs to be found numerically as the potential inside the star is only known numerically. This is done by a binary search in $\mathcal E$, as $\mathcal J_r$ is a decreasing function of $\mathcal E$ at fixed $\mathcal J$, using the circular solution for the given $\mathcal J$ as an upper limit for $\mathcal E$. However, also the integration limits $r_{\min}$ and $r_{\max}$ depend on $\mathcal E$ and have to be computed with every guess of $\mathcal E$. For the orbits of interest here, $r_{\min}$, and sometimes also $r_{\max}$, are less than $R_\star$ and need to be found numerically. This is also done by a binary search since the expression inside the square root in Eq.~(\ref{eq:Jr-star}) is positive for $r_{\min} < r < r_{\max}$ and negative for $r>r_{\max}$ and $r<r_{\min}$. The starting point for this search is the radius  $r_{\rm circ}$ of the circular orbit for the given $\mathcal J$, as one can show that $r_{\min} < r_{\rm circ} < r_{\max}$. The circular radius $r_{\rm circ}$ can be found using $\mathcal J^2=-r_{\rm circ}^3[\ud \Psi(r_{\rm circ}) /\ud r]$, which can be solved efficiently in the Monte-Carlo since the right hand side depends only on the stellar configuration and does not need to be calculated for every WIMP analyzed. To find the orbits of the interesting WIMPs, $\mathcal E$ needs to be determined for all the different stellar stages for which the WIMP orbit crosses the star.

\subsection{Which WIMPs cross the star and when?}
For a WIMP to be captured, it has first to cross the star.  Thus we need to follow how the WIMPs behave in and around the dark star. 

For a WIMP to be captured, it has first to cross the star. 
In the central regions of the halo, the star completely dominates the gravitational potential, as discussed in the previous Subsection. The potential is thus given by $\Psi_\star(r)$, which is Keplerian outside the star and non-Keplerian inside.

Once the star has formed, WIMP orbits not crossing the star are Keplerian. For these orbits, the minimal distance from the center of the star, $r_{\min}$, can be written in closed form as a function of the adiabatic invariants and the mass of the star $M_\star$. Using Eq.~(\ref{eq:EKepler}) to find the smallest zero of the square root in Eq.~(\ref{adiabatic_invariants}) gives for this Keplerian case
\begin{equation}
r_{\min}^{\rm K}=\frac{(\mathcal J_r+2\pi\mathcal J)^2}{4\pi^2GM_\star}\left(1-\sqrt{1-\frac{4\pi^2\mathcal J^2}{(\mathcal J_r+2\pi\mathcal J)^2}}\right)\label{rmin}.
\end{equation}
This radius depends on time through the time dependence of the star mass $M_\star$.

If $r_{\min}^{\rm K}$ is greater than the radius of the star $R_\star$, the orbit will not cross the star at this stage in the stellar evolution. Similarly, if $r_{\min}^{\rm K}$ is smaller than the radius of the star $R_\star$, the orbit will cross the star at this stage in the stellar evolution. One can be sure of these conclusions because  the Keplerian solution is unique, given $\mathcal E$, $\mathcal J$ and the point mass potential. Hence, if the Keplerian orbit solution is found to be partly inside the star, no Keplerian solution with the orbit completely outside the star exists and so the true orbit must have parts inside the star. Conversely if the Keplerian orbit solution is found to be completely outside the star one has found the true orbit since the potential is Keplerian outside the star.

\begin{figure}
\centerline{\epsfig{file=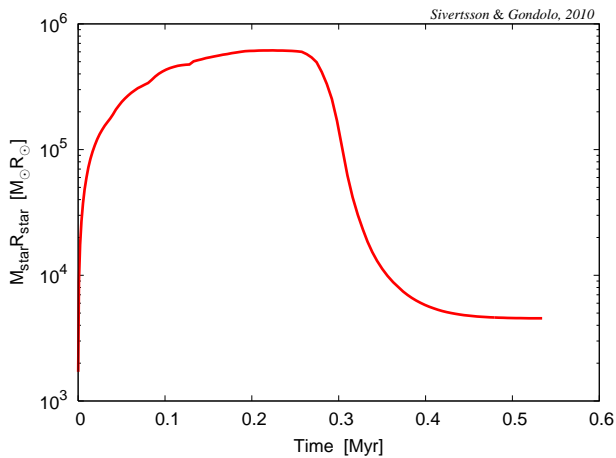,width=0.48\textwidth}}
\caption{The evolution of $M_\star R_\star$ during the star formation. First $M_\star R_\star$ grows as the star grows but then drops fast as the star approaches the main sequence.}
\label{starMR}
\end{figure}

As we are only interested in WIMPs on orbits crossing the star only orbits  fulfilling $r_{\min}^{\rm K}\leq R_\star$ at a given time are interesting. As long as the individual WIMPs do not scatter, the adiabatic invariants in Eq.~(\ref{rmin}) are conserved, implying that $M_\star r_{\min}^{\rm K}$ is constant in time (it is an adiabatic invariant). Along with the requirement that WIMPs crossing the star have to fulfill the relation $M_\star r_{\min}^{\rm K}\leq M_\star R_\star$, one finds that the star can potentially access the maximum amount of dark matter when $M_\star R_\star$ is maximized. Fig.~\ref{starMR} shows the behavior of the quantity $M_\star R_\star$ as a function of time for the Spolyar \textit{et al} canonical case. At first $M_\star R_\star$ increases as the dark star grows, meaning that more and more WIMPs can cross the star. At the time of the Kelvin-Helmholtz (KH) contraction ($t\sim 0.3$ Myr), the quantity $M_\star R_\star$ decreases rapidly, leaving behind WIMPs that were once in stellar contact. For example, WIMPs with $M_\star r_{\rm min}^{\rm K} \sim 10^5 \, M_\odot R_\odot$ cross the star for times between 0.02 Myr and 0.3 Myr, but can no longer cross the star or be captured after 0.3 Myr.
This makes it hard for the star to capture WIMPs in the later stages after the KH contraction. As a consequence, the accessible amount of dark matter that has survived the formation of the star might not be high enough to support the dark star after its KH contraction. 

\subsection{The fate of the WIMPs}

Once we know which Monte-Carlo WIMPs cross the star at a given stage of the dark star evolution, the next step is to find their orbits. Some of these orbits are only partially inside the star, some are completely inside the star. Knowing the gravitational potential inside and outside the star and the adiabatic invariants $\mathcal J$ and $\mathcal J_r$, enough information is given to determine $\mathcal E$ from Eq. (\ref{adiabatic_invariants}). This is done as explained at the end of Section II.B. Having determined the energy and the angular momentum of the WIMP in orbit, the full orbit is specified up to the spherical symmetry (that is up to the specification of the plane of the orbit and of the angle of the periilun \footnote{From the Basque word ilun meaning dark.}). 

In the Monte-Carlo, to find the fate of the WIMP we need to calculate the probability per unit time for the WIMP to scatter off nucleons in the star as well as the probability for the WIMP to annihilate with other WIMPs in the star. In the following, we derive the scattering and annihilation rates at a given stage of the stellar evolution.

When on an orbit crossing the star, the WIMP spends some time
\begin{equation}
\ud t= 2 \, \frac{\ud r}{v_r}
\end{equation}
in the shell between $r$ and $r+\ud r$. Here the factor of 2 accounts for the fact that the WIMP can cross the shell twice, once on its way in and once on its way out; moreover,
\begin{equation}
v_r = \sqrt{v^2-({\mathcal J /r})^2} 
\end{equation}
is the (magnitude of the) radial velocity, where
\begin{equation}
v = \sqrt{2[\Psi_\star(r)-\mathcal E]}
\end{equation}
is the speed at position $r$. 
The probability that in one passage through the star the WIMP annihilates against another WIMP in the dark matter population within the star is then given by 
\begin{equation}
\mathcal P_{\rm ann} = 2 \int_{r_{\rm min}}^{\min(r_{\rm max},R_\star)} \sigma_{\rm ann} \, v \, n_{\rm DM}(r) \, \frac{\ud r}{v_r} ,
\end{equation}
where for the number density of WIMPs $n_{\rm DM}(r)$ at radius $r$ we take the results from the dark star evolution in Ref.~\cite{Spolyar:2009nt}.  For WIMPs on orbits completely inside the star, the phrase ``one passage though the star'' refers to one revolution of the WIMP orbit. The number of times a WIMP passes through the star in the time $\Delta t$ is
\begin{equation}
\frac{\Delta t}{T},
\end{equation}
where the orbital time $T$ is determined numerically from the integral
\begin{equation}
T = 2 \int_{r_{\rm min}}^{r_{\rm max}} \frac{\ud r}{v_r} .
\end{equation}
Finally the probability that a WIMP annihilates during the time interval $\Delta t$ is
\begin{equation}
P_{\rm ann} = \frac{\Delta t}{T} \, \mathcal P_{\rm ann} . \label{annprob}
\end{equation}
We annihilate Monte-Carlo WIMPs according to this probability.

When passing the stellar material there is also some probability for the WIMP to scatter off nuclei in the star. Here only scattering off hydrogen is considered since we assume the scattering cross section is spin-independent, and so no scattering occurs off helium. Heavier elements are a negligible component of the material forming the first stars. We assume that at all radii hydrogen contributes 75\% of the stellar mass density.

In one passage through the star the probability for the WIMP to scatter in a shell at radius $r$ is given by
\begin{equation}
\ud \mathcal P_{\rm scatt} = 2\,\sigma_{\rm scatt}\, v \, n_{\rm H}(r) \, \frac{\ud r}{v_r} .
\end{equation}
Here the factor of 2 accounts for the the fact that the WIMP crosses the shell twice, and $n_{\rm H}(r)$ is the number density of hydrogen atoms in the dark star (taken from Ref.~\cite{Spolyar:2009nt}). The scattering probability per WIMP passage through the star is then given by
\begin{equation}
\mathcal P_{\rm scatt} = \int_{r_{\rm min}}^{\min(r_{\rm max},R_\star)} \ud P_{\rm scatt} .
\end{equation}
Then the probability that a WIMP scatters in the time $\Delta t$ is
\begin{equation}
P_{\rm scatt} = \frac{\Delta t}{T} \, \mathcal P_{\rm scatt} .\label{scattprob}
\end{equation}

In a scattering event, the WIMP loses an amont of energy $\Delta\mathcal E$ subject to the condition
\begin{equation}
\Delta\mathcal E\leq \left(\frac{m_\chi-m_H}{m_\chi+m_H}\right)^2 [\Psi(r)-\mathcal E] .
\end{equation}
Here $\Psi(r)-\mathcal E$ is the WIMP kinetic energy before the scattering at the scatterer location, and $m_H$ is the mass of the target nucleus, which in this paper is always hydrogen since we assume the spin dependent scattering cross section to dominate. All possible values of the energy loss $\Delta E$ are equally probable, making it straightforward for us to determine the WIMP energy $\mathcal E'$ after the scattering.

At a given evolutionary stage of the star, we use the scattering probability per unit time in Eq.~(\ref{scattprob}) to decide how long it takes before the WIMP scatters. Then we compare this time to the time left in the given evolutionary stage to decide if the WIMP scatters before the dark star evolves to a new stage. Afterwards, the Monte-Carlo uses the annihilation probability per unit time in Eq.~(\ref{annprob}) to evaluate if the WIMP annihilates before it scatters or before the current evolutionary stage ends. If the WIMP scatters, the Monte-Carlo uses the distribution $\ud \mathcal P_{\rm scatt}/\mathcal P_{\rm scatt}$ to determine in which shell the WIMP scatters. In the scattering, the WIMP loses energy and ends up on a lower energy orbit, specified by new values of $\mathcal E'$ and $\mathcal J'$. The new direction of the scattered WIMP, and hence $\mathcal J'$, are easily determined as the scattering is isotropic in the center of mass system. After the WIMP is scattered, it is allowed to scatter again in the time left before the next evolutionary stage of the dark star.

When passing to the next evolutionary stage, the adiabatic invariant $\mathcal J'_r$ is needed and can be determined using Eq.~(\ref{adiabatic_invariants}) with the $\mathcal E'$ and $\mathcal J'$ of the last orbit of the scattered WIMP. This allows us to follow the WIMP response to the stellar evolution whether the WIMP scatters or not. If the WIMP scatters many times it eventually sinks towards the core of the star and annihilates with the central WIMP population, injecting its energy into the dark star.

Ref.~\cite{Sivertsson:2009nx} contains further details and discussions on the scattering process. There one can also find how to determine the angular momentum of the WIMP after it has scattered.

\section{Results}
\label{sec:results}

\begin{figure}
\centerline{\epsfig{file=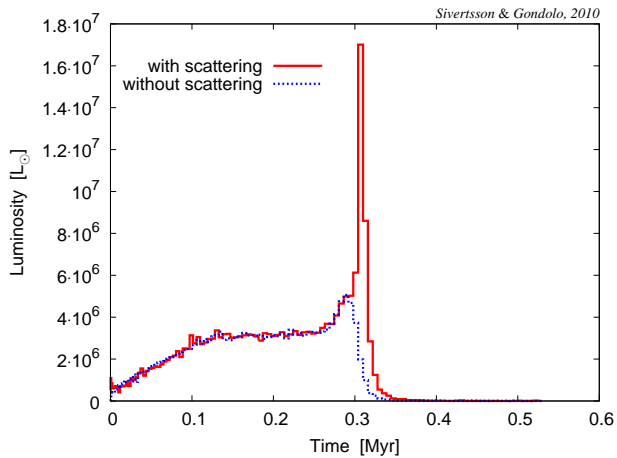,width=0.48\textwidth}}
\caption{The time evolution of the dark star luminosity from WIMP-WIMP annihilations. The top solid (red) line includes WIMP-nucleon scattering in addition to annihilation; the bottom dashed (blue) line does not include scattering.}
\label{luminosity}
\end{figure}

Our Monte-Carlo follows a high number of WIMPs from the initial NFW halo until they annihilate or the star has formed. It allows us to track how much energy dark matter deposits into the star at the different stages of stellar evolution. Fig.~\ref{luminosity} shows the resulting annihilation luminosities with and without scattering. The top solid (red)  line includes WIMP-nucleon scattering, while the bottom dashed (blue) line does not.
 At the early stages of star formation, scattering of WIMPs off star nuclei is not important because the baryon density inside the dark star is low. As the dark star becomes denser, scattering becomes important, provided the scattering cross section is large enough. For the stellar evolution in \cite{Spolyar:2009nt}, the dark star becomes dense enough for scattering to be important just as it starts its rapid KH contraction onto the main sequence (see Fig.~\ref{starMR}). With the rapidly increasing stellar density at this stag,e capture becomes very efficient and via repeated scattering the captured WIMPs rapidly sink toward the center of the star where the dark matter density is high. This makes the captured WIMPs annihilate fast and gives the burst in luminosity seen in Fig.~\ref{luminosity}. As long as capture is efficient, the high central WIMP density is rapidly refilled with more captured WIMPs.

However, the burst lasts a short time because (1) the KH contraction is rapid and the volume of dark matter available for annihilation decreases quickly, and (2) it takes a short time to use up all the WIMPs that can be efficiently captured by the dark star, and no other WIMPs are available for efficient capture. The latter reason (``death by starvation'') is the most important one, as discussed in connection with Fig.~\ref{luminositylong74} in Section~\ref{sec:discussion}. The star is very fast at eating the WIMPs that it can eat easily.  After the contraction, WIMP annihilation continues but at a much smaller rate; in the case we study, the annihilation after contraction is suppressed by (very roughly) about 15 orders of magnitude (five orders of magnitude coming from the change in stellar volume and two times five orders of magnitude coming from the decrease in density due to annihilation and scattering, extracted from Fig.~\ref{density90} below).

When WIMPs scatter, they lose energy, go on smaller orbits, and spend more time near the center of the star where the density is higher and the annihilation faster. 
\begin{figure}
\centerline{\epsfig{file=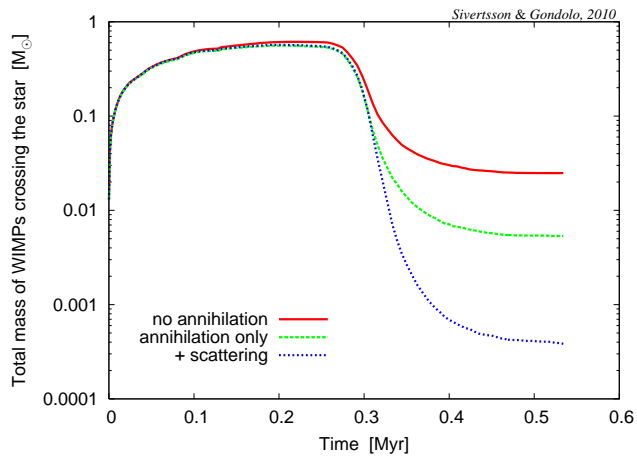,width=0.48\textwidth}}
\caption{The dark matter mass accessible to the star as a function of time. In other words, the total mass of the WIMPs on orbits intersecting the star at the different stellar evolution stages. The dotted (blue) line includes both WIMP scattering and annihilation. The dashed (green) line includes WIMP annihilation but no scattering. The solid (red) line includes neither scattering nor annihilation.}
\label{alive}
\end{figure}
Both annihilation and scattering remove WIMPs from orbits that cross the star. This is illustrated in Fig.~\ref{alive}, where the mass of dark matter accessible to the star (i.e.\ the total mass of WIMPs on orbits crossing the star) is plotted as a function of time. The three lines illustrate the effect of the stellar evolution as well as that of scattering and annihilation. The top (solid red) line is similar to the curve in Fig.~\ref{starMR}, and includes just the gravitational response of the WIMPs to the star and neither scattering nor annihilation. It shows that the mere contraction of the star already makes it a much smaller target for WIMP capture, by about one order of magnitude in the accessible mass. Annihilation reduces the number of WIMPs that cross the star (green dotted line), as annihilation removes WIMPs from the system. The inclusion of scattering reduces this number further, which might at first not seem obvious. The reason is that WIMPs on orbits not crossing the star cannot scatter in the star. Hence the total mass of WIMPs on orbits crossing the star can never increase due to scattering. Furthermore, scattering puts WIMPs onto lower energy orbits, making them more vulnerable to annihilation. Thus the number of WIMPs in stellar contact is further reduced (see Fig.~\ref{alive}). As discussed in Section~\ref{sec:discussion}, our Monte-Carlo only includes WIMPs from the inner 1 percent of the NFW halo scale radius $r_s$, but WIMPs from regions further out do not contribute significantly to the annihilation rate inside the dark star.

\begin{figure}
\centerline{\epsfig{file=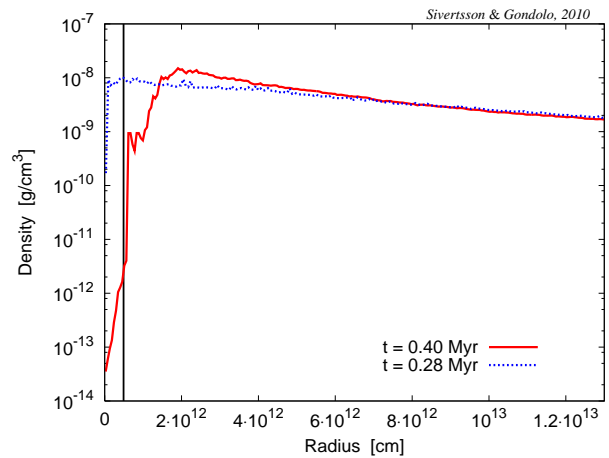,width=0.48\textwidth}}
\caption{The density profile of dark matter around the star just before (dotted blue line) and just after (solid red line) the Kelvin-Helmholtz contraction at the end of the dark star phase  (scattering is included when generating this graph). The vertical straight line marks the radius of the star after contraction; before contraction, the radius of the star is 4.2$\times10^{13}$ cm, about four times the length of the horizontal axis.}
\label{density90}
\end{figure}

From the Monte-Carlo one can also extract the dark matter density profile around the formed star. This is shown in Fig.~\ref{density90} for two different times, just before and just after the KH contraction (blue dotted line and red solid line, respectively).  The solid red line is for the time 0.4~Myr, which is just after nuclear burning has started. The blue dotted line shows the time 0.28~Myr, which is just before the star contracts enough for scattering to be important, as can be seen in Fig.~\ref{luminosity}. The vertical, black line marks the size of the star at $t=0.4$~Myr. At $t=0.28$~Myr the stellar radius is $4.2\times 10^{13}$~cm and hence outside the graph. The unevenness of the graph comes from having simulated a finite number of WIMPs; the very central region has far less simulated WIMPs (as the fraction of WIMPs reaching so far in is very low) and hence greater inaccuracy.

The KH contraction of the dark star increases the dark matter density in the central region through adiabatic contraction. The density peak at $\sim 2\times10^{12}$ cm on the solid line in Fig.~\ref{density90} is a residual of this density increase. Moreover, scattering increases the density at the very center by moving WIMPs into low energy orbits on which they spend more time near the center. The underdense ``hole'' at small radii on the solid line in Fig.~\ref{density90} is a consequence of the rapid WIMP scattering and annihilation. WIMPs with large contributions to the dark matter density inside the dark star are very vulnerable to scattering which makes them sink to the core and annihilate away fast.
In the absence of inhomogeneities, this final configuration is stationary: WIMPs in orbits that crossed the star have scattered and annihilated away, and WIMPs in orbits that do not cross the star remain in those orbits. Even if the star would move in space, the WIMPs surrounding it would move with it, much like the Moon moves with the Earth while orbiting it. As long as this setting is not disturbed by inhomogeneities, the WIMP density profile around the star will look essentially the same until the star dies (except that the annihilation timescale near the density peak is short and annihilation will reduce the density). The star will then simply be a normal massive star, living its life undisturbed by its envelope of dark matter. 

\section{Discussion}
\label{sec:discussion}

We mentioned in Section \ref{sec:results} that the death of a dark star is due to dark matter exhaustion rather than rapid KH contraction. In the first case, the WIMP supply runs out first and the lack of luminosity from WIMP annihilation causes the dark star to contract onto the main sequence. In the second case, the dark star contracts first, for example because the WIMPs do not provide enough luminosity, and is then unable to continue capturing enough WIMPs to remain a dark star because it has become too small. There is a third possibility, namely that the dark star does not contract or might even grow, since the extra luminosity provided by the captured WIMPs make it puffier and it may become more massive by accreting gas for a longer time. If the dark star does not grow, there is a maximum amount of WIMP mass that can be captured before going onto the main sequence, given by the graph in Fig.~\ref{alive} at the time just before contraction $\sim 0.27$ Myr. 

\begin{figure}
\centerline{\epsfig{file=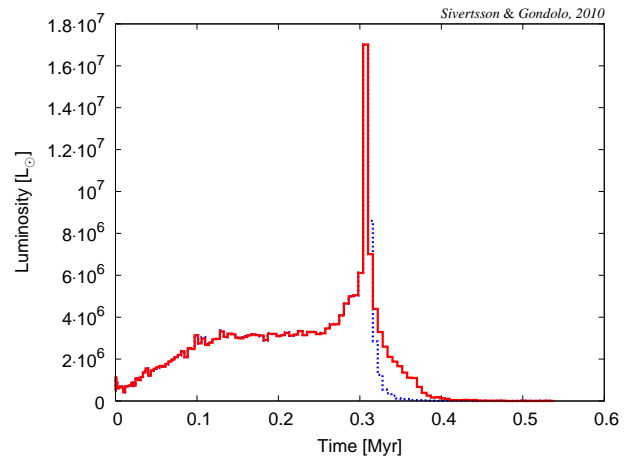,width=0.48\textwidth}}
\caption{The energy injected into a dark star from WIMP annihilation for a star with modified stellar evolution so that it is artificially frozen just before contraction. The WIMP luminosity in the altered stellar evolution is given by the solid (red) line; for reference, the WIMP luminosity without altering the stellar evolution is shown by the dashed (blue) line. Both luminosities include scattering. This is to illustrate that the disappearance of dark matter in contact with the star is due to the depletion of star-crossing orbits with low angular momentum and is not an artifact of analyzing a star that contracts too rapidly.}
\label{luminositylong74}
\end{figure}

Here we argue that even if the dark star does not contract, the supply of dark matter is rapidly exhausted (the first case above). In principle, we should recompute the evolution of the dark star in the light of the capture results we find, instead of using the dark star evolution tracks without capture in \cite{Spolyar:2009nt} (notice that we cannot use the evolution tracks with capture in \cite{Spolyar:2009nt} because they are based on Gould's capture formula that does not apply to our case). A detailed numerical study of the response of the dark star to the extra luminosity produced by captured WIMPs is under way \cite{myself}.  Here we avoid going into the details of the stellar response; we instead artificially freeze the evolution of the dark star on the verge of contraction. We prolong the stage in the stellar evolution when the dark matter capture via scattering is at a maximum (the peak in Fig.~\ref{luminosity}) from 6,000 to 60,000 yr. The dark matter luminosity for the star with this slight change in the stellar evolution is shown in Fig.~\ref{luminositylong74}, which is analogous to Fig.~\ref{luminosity} but for the KH contraction which is artificially delayed as just explained. We still  observe a drastic reduction of the luminosity due to WIMP annihilation. We find that the extra dark matter supplied via WIMP nucleon scattering can only support the star for a limited period of time; it does not result in a long lasting dark star phase. In other words, we conclude that the demise of the dark star phase is due to the exhaustion of the dark matter supply. 

In our code we have not included WIMP annihilations that occur outside the star. We expect these annihilations are not important for the energy production in the dark star for the following reason. The dark star cannot absorb the energy released by WIMP annihilations outside of it, but the outside WIMP annihilations would somewhat reduce the number of WIMPs available for annihilation inside the dark star. However, during the dark star phase only a very small fraction of the accessible WIMP population would have annihilated away outside the dark star, making this depleting effect unimportant for the dark matter heating of the dark star. Dark matter annihilations can also affect the cooling of the dark star by ionizing the gas (discussed in \cite{Ripamonti:2010ab}). We are here only looking at the dark matter component of the dark star process so feedback through gas ionization is beyond the scope of our paper.

The conclusion that the main sequence star does not significantly benefit from its surrounding dark matter halo is based on the assumption that the final dark matter configuration, as shown in Fig.~\ref{density90}, is stable enough against perturbations to largely sustain its shape over time. For example a close encounter with another star could severely affect the dark matter envelope. However, without long term heating from WIMP annihilations, stars this massive are very short lived, with expected lifetimes of less than 2~Myr \cite{Schaerer:2001jc}. Since we assume there is only one dark star per dark halo, the time scale for the dark star to encounter another star is expected to be much longer than the star's lifetime. 

The star locally completely dominates the gravitational potential so if the star experiences a small acceleration towards some distant object the dark matter envelope would just follow, keeping its shape and distance to the star. If the star experiences a more violent event, so that it is suddenly kicked out of its position and thrown into its dark matter envelope, the star will once again be able to benefit from the dark matter. However, the system will fast reconfigure itself around the new gravitational center at the stellar location. The star would quickly eat up the WIMPs it can eat easily, and the other WIMPs will have too high an angular momentum relative to the new gravitational center to be reachable by the star. Such a drastic event would hence induce a short dark matter burst similar to the capture burst in Fig.~\ref{luminosity}, followed once again by starvation of the star. For the star to have a long-term benefit from the dark matter envelope, continuous perturbations of the star or the dark matter envelope are needed.

The initial star-forming gas cloud could also have some initial velocity relative to the dark matter halo. The forming star will still create a dark matter envelope around it. The star dominates the gravitational potential and carries the WIMPs with it, hence the WIMPs in the star's dark matter envelope do not feel that the halo has a net velocity relative to the star. In brief, the initial conditions of the WIMP orbits in the dark matter envelope will be slightly different but the dynamics of the system will be the same, so the final result for the dark matter in the central region of the halo will essentially be the same. The only WIMPs that will notice the relative velocity between the star and the halo are the WIMPs traveling far enough out in the halo to also probe the halo's gravitational potential. However, as these WIMPs pass the region where the star is, their velocities are very high, making any reasonable stellar velocity negligible in comparison. We also show below that these WIMPs are not important anyway. Finally, a net velocity of the star-forming cloud could have a significant impact only if the star would form far away from the center of the halo and thus miss the halo central high dark matter density. As long as the formation is not very far from the center, the off-center location would primarily reduce the initial dark matter heating from adiabatic contraction as this contraction starts before the star has taken its dominance over the local gravitational potential.

The star's host dark matter halo is expected to have substructure with smaller subhalos of dark matter. The effect of these subhalos on the formation of dark stars is still to be explored. In any case, subhalos in the central region are likely to be destroyed by the formation of the dark star. Since subhalos are only a fraction of the total dark matter population around the star, and since each subhalo would provide WIMPs for a limited amount of time and there would be gaps in the WIMP supply between subhalo crossings, the subhalos are not expected to sustain a dark star through capture over a long period of time. It could be possible, however, that large- and small-angle gravitational scattering of WIMPs off subhalos around the dark star changes the phase-space structure of the dark matter envelope. A similar effect would arise from inhomogeneities in the gas surrounding the dark star. Many small-angle deflections could over time make some WIMPs in the envelope end up crossing the star. 

The part of the dark matter envelope closest to the star should be the most vulnerable to perturbations that make the WIMPs cross the star. If the perturbations would be such that the star first accesses the WIMPs in its immediate surroundings, it would need to capture all of the dark matter out to a radius of $10^4 R_\star$ (after KH contraction) to prolong its lifetime by roughly 1 Myr. Hence, even if the dark star could efficiently capture the dark matter in this closest reservoir, it is still not enough to substantially prolong the life of the star. To do so, the dark star needs to efficiently access dark matter from outside $10^4 R_\star$. By a simple geometrical argument, WIMPs further away from the star should be less likely to be perturbed by inhomogeneities into orbits crossing the star. Orbits stretching further out also have longer orbital times, giving fewer stellar crossings per unit time. Assuming that all WIMPs are constantly subject to perturbations, and using a simple geometrical probability factor $(R_\star/r)^2$ for the probability per orbital revolution that an orbit extending out to radius $r$ intersects the star, gives an estimate of the capture rate and of the luminosity contribution from WIMPs on orbits stretching further out than $10^4 R_\star$. This luminosity is too small to sustain the dark star phase.
 
The overall conclusion of this is that it could in principle be possible for WIMPs from the dark star's WIMP envelope to temporarily feed the main sequence star with enough dark matter to relaunch a short dark star phase, for example through perturbations of or direct interaction with dark matter subhalos in the dark matter halo. What the above argument shows is however that it is very unlikely for the dark matter envelope to be able to continuously supply a dark matter capture rate high enough to obtain a dark star phase comparable or longer than the star's main sequence lifetime.

In the tradition of previous papers about Pop III dark stars, this work assumes that the first stars form individually near the center of a dark matter halo. Some recent work \cite{Turk:2009ae} has suggested that some of the first stars might have formed as binary systems. If that would be the case, the analysis of the dark matter response during their formation would be severely more complicated. In general, one would expect a less important density increase due to adiabatic contraction and also, once the nearby star has formed, their motion should accelerate and throw out WIMPs bound to the central region, further reducing the central density. To make this statements more precise, a specific investigation is needed.

Population~III stars as massive as discussed here can become unstable to pulsations \cite{Baraffe:2000dp,Schaerer:2001jc}. An interesting thought is that as the star pulsates it has access to a part of its dark matter envelope. This could potentially have interesting consequences, but would require a study in itself. In any case, the dark star can only eat up the dark matter in the envelope during the first few pulsations, since in subsequent pulsations the dark matter supply will have been depleted. Moreover, the gas density in the outer layers of the pulsating star may be too small for scattering of WIMPs to occur with any significance.

To speed up the Monte-Carlo calculation, when picking WIMPs from the initial NFW halo, we restrict ourselves to the region $r<0.01r_s$. In principle, one should also consider WIMPs further out, but the probability that they scatter in the dark star decreases rapidly with distance and they soon become irrelevant. In addition, the orbital time for WIMPs with orbits extending out to $0.02 r_s$ is $\sim1$ Myr, meaning that these WIMPs will not be captured by the dark star within the much shorter lifetimes we find. We have checked that WIMPs beyond 0.01$r_s$ are indeed irrelevant for our results by running a Monte-Carlo simulation picking WIMPs from a region that extended tenfold in radius, $r<0.1r_s$.  

In our derivation, we have neglected the contribution from WIMPs that are not bound to the star. Here we present several arguments leading to the conclusion that the maximum rate at which WIMPs can be captured by a dark star from the unbound WIMP population is indeed small enough to be completely negligible. Unbound WIMPs are on orbits that extend beyond the radius of influence of the star $r_{\rm infl}$. This radius was estimated in Section II.B to be $r_{\rm infl}\sim 0.01 r_s$ for the capturing stages of a dark star, namely the final stages with $M_\star\sim10^3\,M_\odot$ and $R_\star=4\times10^{11}\,{\rm cm}$. Now, the density of WIMPs at $r\sim r_{\rm infl}$ is comparable to the original density in the NFW halo, namely $\sim 10^3 \, {\rm GeV/cm^3}$. This density is much smaller than the density of bound WIMPs near the star, which is of order $10^{11} \, {\rm GeV/cm^3}$. Therefore unbound WIMPs do not contribute significantly to the WIMP density near the star. One might worry that gravitational focusing would be able to collect more unbound WIMPs than the previous argument would indicate. However, the focusing effect can be included in an estimate of the amount of dark matter mass that could fall in from a distance of $r_{\rm infl}$ and cross the star. One finds for this rate of mass collection $\dot{M}_{\rm unbound} \sim \pi R_\star^2 F_\star$, where $F_\star = F(r_{\rm infl}) v_{\rm esc}^2/v(r_{\rm infl})^2$ is the inward flux of unbound WIMPs at the surface of the star, $v_{\rm esc}^2/v(r_{\rm infl})^2$ is the gravitational focusing factor, and $F(r_{\rm infl})=\rho(r_{\rm infl}) v(r_{\rm infl})$ is the inward flux at $r=r_{\rm infl}$. Estimating the density and speed of WIMPs at $r=r_{\rm infl}$ using $\rho_{\rm NFW} \propto r^{-1}$ and $v(r_{\rm infl}) = [GM_{\rm NFW}(r_{\rm infl})/r_{\rm infl}]^{1/2}\sim 1$ km/s, gives a mass collection rate of $\sim 10^{-5} \, M_\odot/{\rm Myr}$ from the population of unbound WIMPs. To survive at a luminosity of $\sim 10^7\, L_\odot$, the star would need an inflow of dark matter WIMPs to annihilate equal to $\sim 1 M_\odot/{\rm Myr}$, much higher than what unbound WIMPs could provide. Also, only a small fraction of the WIMPs passing through the star will actually scatter and be captured by the star, reducing the heating from unbound WIMPs further. In addition, the orbital periods of WIMPs on unbound orbits, i.e.\ orbits that reach out to $r=r_{\rm infl}$ or more, is at least 2 Myr (assuming Keplerian orbits in this estimate). This time is comparable to the lifetime of the dark star, including the later main sequence stage. Thus unbound WIMPs cannot reach the star in time to keep it alive through dark matter annihilation.

It has recently been argued in \cite{Freese:2010re} that triaxiality of the dark matter halo could efficiently continuously supply WIMPs to the star and maintain the dark star phase on cosmological time scales. However, as the star dominates the gravitational potential in the central part of the halo, only WIMPs traveling out to radii comparable to the star's radius of influence $r_{\rm infl}$ feel the triaxiality of the dark matter halo. Thus the method in the previous paragraph can be used to estimate the contribution to the capture rate from WIMPs probing the triaxiality of the halo. This estimate gives that the supermassive dark star's capture rate from the triaxial halo in \cite{Freese:2010re} is too low by a factor of $\sim 10^9$ to sustain the luminosity of the star. Even though this estimate is based on the outer halo being isotropic, accounting for anisotropies or chaotic orbits is not expected to increase the capture rate by a factor of $\sim 10^9$. A proper treatment of the triaxiality may slightly increase the capture rate thanks to elongated orbits that are bound to the star and almost reach $r_{\rm infl}$, but again the increase cannot be by 9 orders of magnitude. In summary, a triaxial halo could feed the star with dark matter for a very long time but as the capture rate is too low to sustain the dark star, triaxiality of the halo is not interesting for the capture rate.

An intriguing possibility is that long-lived dark stars may form from the ashes of Pop III dark stars (Phoenix Dark Stars). When the main sequence star following the dark star phase explodes, it ejects gas into a region that can still have a  considerable density of dark matter WIMPs. This dark matter is the remnant of the dark matter envelope that surrounded the dark star. In fact, for $\sigma_{\rm ann} v = 3\times 10^{-26}\,{\rm cm^3/s}$, a remnant dark matter density $ n_0 = 1/(\sigma_{\rm ann} v t_0) \sim 10^8 \, {\rm WIMPs/cm^3}$ can be sustained for a time $t_0\sim10 \, {\rm Gyr}$ (approximately the age of the universe). For a 100 GeV WIMP, this corresponds to a mass density $\rho_0 = 10^{10} \, {\rm GeV/cm^3} = 2\times 10^{-14}\, {\rm g/cm^3}$. Some of the ejected gas may fragment and form Pop II and later Pop I stars, which can move relative to the dark matter remnant and capture WIMPs from the environment in the way described by previous authors \cite{Yoon:2008km,Taoso:2008kw,Freese:2008ur}. These ``daughter'' stars are expected to be less massive than their parent Pop III dark stars, and thus would need less dark matter to be a dark star. Given appropriate environmental conditions, some of these second generation dark stars might perhaps survive until today.

\section{Conclusions}

As found by previous authors, the density of dark matter inside a Pop III protostar can be high enough to lead to a dark star phase powered by dark matter annihilation instead of nuclear fusion. A large WIMP-nucleon scattering cross section may further increase the energy injected into the dark star by WIMP annihilations. As discussed in the introduction, many authors have claimed that the WIMP supply from scattering may last a long time, even until the present day. Our work does not support those findings. 

We find that WIMP capture is not important until the dark star contracts on its way to the main sequence and becomes dense enough to be a viable dark matter target. We also find that WIMP capture lasts a relatively short amount of time as the WIMPs available for capture are quickly depleted in less than $10^5$ years and their orbits are not replenished.

It is also very unlikely that perturbations in the star's dark matter envelope could keep the inflow of dark matter high enough to support a further prolongation of the dark star phase, namely for lifetimes much longer than the $10^5$ years we here calculated. Furthermore the inflow of dark matter from the outer regions of the halo is too small to be of any interest. From this we conclude that a single Pop III star forming at the center of an NFW halo will have a total lifetime of order a few million years and that the dark star phase supported by WIMP capture will not substantially prolong the lifetime of the dark star.

Our work indicates that WIMP capture in Pop III dark stars is not able to make them live forever. However, later generations of less massive stars produced in the dark matter environment left behind by Pop III dark stars may capture enough dark matter to form a second generation of dark stars.

\acknowledgments

We thank Pat Scott and Joakim Edsj{\"o} for very rewarding discussions. In particular, we are grateful to Douglas Spolyar, Peter Bodenheimer, and Katherine Freese for letting us use an unpublished set of dark star evolution models from \cite{Spolyar:2009nt}. This work was partially supported by the National Science Foundation under award no.\ PHY-0456825 (P.G.), and by the Swedish Research Council (Vetenskapsr{\aa}det) under contract no.\ 315-2004-6519 (S.S.). We thank the Bakficken Cafe, Visby, where this work was completed.

\bibliography{darkstars}
\end{document}